\def\vector#1{\mbox{$\bm#1$}}
\newcommand{\Tabref}[1]{table~\ref{#1}}
\newcommand{\Figref}[1]{figure~\ref{#1}}

\RequirePackage{fix-cm}
\documentclass[smallextended]{svjour3}       
\smartqed  
\usepackage{graphicx}
\usepackage[top=30truemm,bottom=30truemm,left=25truemm,right=25truemm]{geometry}
\usepackage{color}
\usepackage{amsmath}
\usepackage{lscape}
\usepackage{amsfonts}
\usepackage{bm}
\usepackage{color}
\usepackage{here}
\graphicspath{{./}}

\begin{document}

\title{Machine-learning-based reduced order modeling for unsteady flows around bluff bodies of various shapes}

\titlerunning{Machine-learning-based reduced order modeling for flows around bluff bodies}        

\author{Kazuto Hasegawa        \\
        Kai Fukami   \\
        Takaaki Murata  \\
        Koji Fukagata
}

\institute{Kazuto Hasegawa \at
              Department of Mechanical Engineering, Keio University, Yokohama 223-8522, Japan\\
              Dipartimento di Scienze e Tecnologie Aerospaziali, Politecnico di Milano, Milano 20156, Italy
              \\
           Kai Fukami, Takaaki Murata, Koji Fukagata \at
              Department of Mechanical Engineering, Keio University, Yokohama 223-8522, Japan
              \\
              \email{fukagata@mech.keio.ac.jp}
}

\date{Received: \today / Accepted: ???}

\maketitle

\begin{abstract}
We propose a method to construct a reduced order model with machine learning for unsteady flows.   
The present machine-learned reduced order model (ML-ROM) is constructed by combining a convolutional neural network autoencoder (CNN-AE) and a long short-term memory (LSTM), which are trained in a sequential manner.  
First, the CNN-AE is trained using direct numerical simulation (DNS) data so as to map the high-dimensional flow data into low-dimensional latent space.  
Then, the LSTM is utilized to establish a temporal prediction system for the low-dimensionalized vectors obtained by CNN-AE.  
As a test case, we consider flows around a bluff body whose shape is defined {using a combination of trigonometric functions with random amplitudes.}
{The present ML-ROMs are trained on a set of 80 bluff body shapes and tested on a different set of 20 bluff body shapes not used for training, with both training and test shapes chosen from the same random distribution.}
The flow fields are confirmed to be well reproduced by the present ML-ROM in terms of various statistics.
We also focus on the influence of two main parameters: (1) the {latent vector size} in the CNN-AE, and (2) the time step size between the mapped vectors used for the LSTM. 
The present results show that {the ML-ROM works} well even for unseen shapes of bluff bodies when these parameters are properly chosen, {which implies great potential for the present type of ML-ROM to be applied to more complex flows.}

\end{abstract}

\section{Introduction}

In recent years, a huge amount of detailed flow field information has been accumulated as {\it fluid big data} thanks to high-resolution numerical simulations and image-based measurements.  
Understanding essential phenomena and controlling flows based on such big data --- as they are --- are difficult due to their complexity.
Therefore, reduced order models (ROMs) have been utilized as one way to tackle such problems.  
One of the beauties of ROMs is that they can map a flow field with high dimensions into a low-dimensional space \cite{TBDRCMSGTU2017}.  
Lumley \cite{Lumley1967} introduced the proper orthogonal decomposition (POD), which can express a flow field with several principal modes and the corresponding eigenvalues.  
Schmid \cite{Schmid2010} proposed the dynamic mode decomposition (DMD) that extracts the information from flow fields by focusing on a specific frequency. 
These ROMs are considered to have deepened our understanding \cite{BL1967} and enabled us to control flow phenomena at low computational costs \cite{GB2017}. 
{Despite the great advantages of these linear methods, an annoying problem may be that the number of modes required to represent a flow often becomes too large to handle because nonlinear phenomena must be approximated by a linear superposition of orthogonal modes.
Even for a turbulent channel flow at a low Reynolds number, for instance, 7260 POD modes are required to reconstruct 95\% of its total energy~\cite{AP2007}.
In order to reduce the number of modes, use of the novel nonlinear dimension reduction technique that brought innovation to image recognition \cite{Hinton2006}, i.e., machine learning, can be considered as a good candidate.}

In recent years, machine learning techniques, which can automatically extract key features from tremendous {amount of} data, have achieved noteworthy results in various fields including fluid dynamics owing to the advances in the algorithms centering on {\it deep learning} \cite{LBH2015,Kutz2017,BNK2019,THBSDBDY2019,BEF2019,FFT2020}, which has been enabled by the recent development of {computational power}.  
For instance, Ling et al. \cite{LKT2016} proposed a tensor basis neural network to predict the Reynolds stress anisotropy tensor for Reynolds-averaged Navier--Stokes simulations.  
The proposed method was applied to the duct and wavy flows and it showed substantial merits over the conventional eddy viscosity models.
Fukami et al. \cite{FFT2019} utilized convolutional neural network{s} (CNN) \cite{LBBH1998} for a super-resolution reconstruction of two-dimensional turbulence and reported that the customized CNN model can recover the maximum wavenumbers of energy spectrum from grossly coarse low-resolution flow data.
A machine learning method was also applied to the flow around a circular cylinder so as to predict the flow fields at various Reynolds numbers from the pressure drag coefficient distribution \cite{XPWH2018}.  
Moreover, Viquerat and Hachem \cite{VH2019} have proposed a CNN based method to predict drag coefficients in a two-dimensional low Reynolds number flow around various random shapes generated by B\'ezier curves.
{In this way, capability of machine learning has been demonstrated for different kinds of fluid dynamics problems, although it should be noted that the literature on this topic is vast and many other applications exist despite the references provided here.}

Of particular interest concerning the machine learning for fluid dynamics is its applications to { nonlinear} reduced order modeling.  
San \& Maulik \cite{SM2018} proposed an ROM for quasistationary geophysical turbulent flows based on the extreme learning machine.  
Srinivasan et al. \cite{SGASR2019} proposed a machine learning model based on a multi-layer perceptron and a long short-term memory (LSTM) \cite{HS1997} to successfully predict temporal behaviors of the coefficients in the nine-equation turbulent flow model.  
More recently, Murata et al. \cite{MFF2020} have proposed nonlinear mode decomposition via CNN autoencoder (CNN-AE) and reported its great advantage over POD for the flow around a circular cylinder and its transient process in terms of the feature extraction of flow fields in lower dimensions.  

The objective of the present study is to propose a method of reduced order modeling using CNN-AE and LSTM, which have been separately shown to have great potentials as introduced above.
The machine-learned reduced order model (ML-ROM) proposed here is constructed by combining a CNN-AE and an LSTM, which are trained in a sequential manner. 
The CNN-AE part is trained first to map the high-dimensional flow field obtained by direct numerical simulation (DNS) into a low-dimensional latent space.
Then, the LSTM part is trained to predict the temporal evolution of the low-dimensionalized vectors obtained by the CNN-AE.
As a test case, we consider two-dimensional unsteady flows around a bluff body.  
We randomly define the shapes of bluff bodies in order to assess the performance of the present ML-ROM for unseen data.  
Moreover, the effects of the two key parameters are examined to unveil their {influence} on the model performance.

The remainder of the paper is organized as follows.
Section 2 introduces the details {of} the training data and the theory of the machine learning models. 
The results and case studies {on the} prediction of flows around bluff bodies of various shapes are presented and discussed in section 3.
Finally, the concluding remarks are provided in section 4.

\section{Methods}

\subsection{Training data}

Two-dimensional direct numerical simulation (DNS) of flows around various bluff bodies, whose shapes are defined randomly, are performed to obtain the flow fields used for training, validation, and assessment of the ML-ROM.
The governing equations are the incompressible continuity and Navier--Stokes equations, i.e., 
\begin{align}
  &\nabla\cdot \vector{u}=0,\\
  &\frac{\partial\vector{u}}{\partial t}+\nabla\cdot(\vector{u}\vector{u})=-\nabla p+\frac{1}{\rm{Re}_{\it{D}}}\nabla^2\vector{u},
\end{align}
where $\vector{u}{=[u, v]^T}$, $p$, and $t$ denote the velocity, pressure, and time, respectively.  
All variables are made dimensionless by the fluid density $\rho^*$ , the uniform velocity $U_\infty^*$, and the frontal length $D^*$ of the body, where the superscript $*$ represents dimensional variables.
The Reynolds number is set to $\textrm{Re}_D=U_\infty^*D^*/\nu^*=100$, where $\nu^*$ is the kinematic viscosity.  

\begin{figure}
    \begin{center}
        \includegraphics[clip,width=0.65\textwidth ]{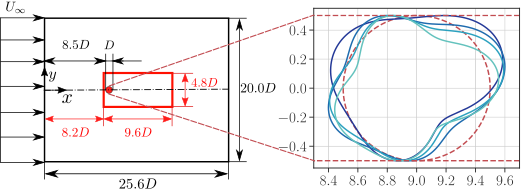}
    \caption{The computational domain used for DNS (black lines), its subdomain used for the machine learning (red lines), and the shapes of the bluff bodies. Blue lines indicate the example of the bluff bodies defined randomly following equations $(3)$ and $(4)$.}
    \label{comp._domain}
    \end{center}
\end{figure}

The computational domain is shown in the left part of \Figref{comp._domain}. 
The center of the bluff body is located $9D$ from the inflow boundary.  
{
The uniform velocity $U_\infty=1$ is given at the inflow boundary, the convective boundary condition is used at the outflow boundary, and the free-slip condition is imposed on the top and bottom boundaries.
}
{

The present DNS code is basically the same as that used by Anzai et al.~\cite{Anzai2017} for flows around a square cylinder, except that a ghost-cell method \cite{KGF2017} is used to satisfy the no-slip boundary condition on the bluff body surface.
The spatial discretization is done by using the energy-conservative second-order finite difference method on a staggered grid system \cite{Amsden1970}, which is uniform in both streamwise ($x$) and transverse ($y$) directions with the grid size $\Delta x = \Delta y = 0.025$. 
The number of computational cells is $(N_x, N_y)=(1024, 800)$.
The time integration is done using the low-storage third-order Runge-Kutta/Crank-Nicolson (RK3/CN) scheme \cite{Spalart1991} with a velocity-pressure coupling similar to the simplified marker and cell (SMAC) method \cite{Amsden1970}.
The time step is set to $\Delta t=2.5\times10^{-3}$.
The pressure Poisson equation is solved by means of the fast Fourier transform (FFT) in
$x$ direction with the mirroring technique \cite{Mitsuishi2007} and the tridiagonal matrix algorithm (TDMA) in ($y$) direction.
We have verified for some selected cases that the present grid resolution is sufficiently fine, and we have validated that the time-averaged drag and rms lift coefficients as well as the Strouhal number computed for a circular cylinder (for which references are available) are in good agreement with the references.}

As mentioned above, the flows around bluff bodies with various shapes are considered in order to examine whether {we can construct a single ML-ROM approximating the function ${\cal F}$ corresponding to the time-discretized Navier--Stokes equation $\bm{q}^{(n+1)\Delta t} = {\cal F}(\bm{q}^{n\Delta t})$ (where $\bm{q}=[u,v,p]^T$ and the superscript denotes time), which is valid even for unseen shapes.}
The shape of a single bluff body is defined as
\begin{align}
    &r=0.5+\sum_{n=1}^4 a_n\sin{n\theta}+\sum_{n=1}^4 a_{n+4}\cos{n\theta},\\
    &\sum_{n=1}^8a_n=0.5,
\end{align}
where $r$ is the distance between the center and the surface, $\theta$ represents the angle from the inflow (i.e., $x$) direction, and $a_n$ denotes random numbers normalized to satisfy equation $(4)$.  
{The bluff body shapes generated using equations (3) and (4) are rescaled so that the frontal length becomes unity and} ${\rm Re}_D=100$ in all cases.  
Fifty different shapes are defined, and the flows around them are produced using the DNS.  
Moreover, the flow fields are rotated around the $x$-axis symmetry to increase the amount of training data.  
In this way, hundred kinds of flows are prepared as the data sets.
{Note in passing that the achievable range of shapes generated using equations $(3)$ and $(4)$ is limited, and the use of this formulation is intended to be a proof of concept.}

\begin{figure}
    \begin{center}
        \includegraphics[clip,width=0.9\textwidth ]{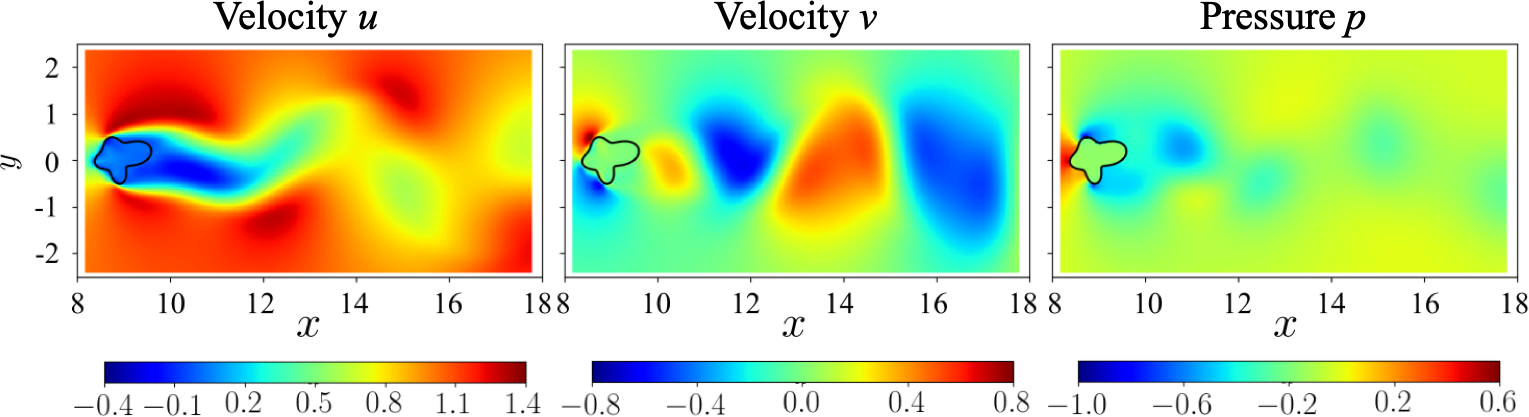}
    \caption{An example of the velocity and pressure fields around the randomly defined bluff body.}
    \label{flow_fields_DNS}
    \end{center}
\end{figure}

In order to focus on the flow around the bluff body, the velocities and pressure $(u, v, p)$ in the region enclosed by the red line in \Figref{comp._domain} are extracted to use for machine learning. 
The size of the instantaneous field data used for ML-ROM construction is $(\hat{N_x}, \hat{N_y}, N_{\phi}) = (384, 192, 3)$, where $\phi$ represents {the considered} physical {quantities}.
An example of the flow fields is shown in \Figref{flow_fields_DNS}.
{In this study, we do not apply any data pre-processing such as normalization or standardization since the order of magnitude is unity for all the quantities thanks to the nondimensionalization, and the bluff body shapes are adjusted to have the equal frontal length (i.e., unity) as mentioned above.}

\subsection{Machine learning}
\subsubsection{Convolutional neural network autoencoder (CNN-AE)}

The convolutional neural network (CNN) \cite{LBBH1998} has been widely used in the field of image recognition, and it has also been applied to fluid dynamics in recent years \cite{FFT2019,XPWH2018,FNKF2019} due to its ability to deal with spatially coherent information.  
The CNN is formed by connecting two kinds of layers: convolution layers and sampling layers.

\begin{figure}[!b]
    \begin{center}
        \includegraphics[clip,width=0.8\textwidth ]{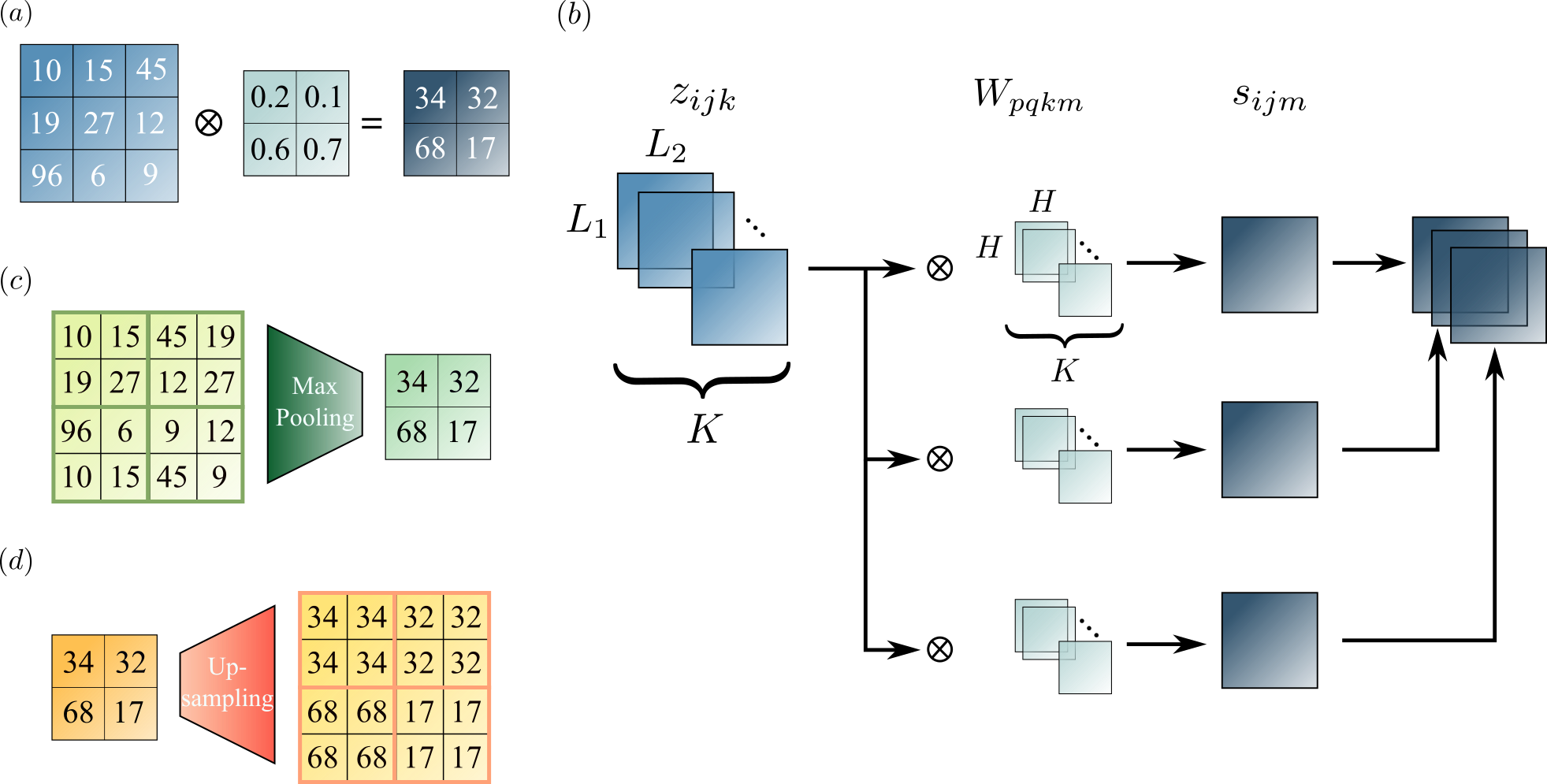}
    \caption{Operations {in} the convolutional layer and the sampling layer{:} $(a)$ {c}onvolutional operation using a weighted filter $W${;} $(b)$  {t}he computation in the convolution layer {with} $M=3${;} 
     $(c)$ {m}ax pooling operation. {$(d)$ {u}psampling operation.}}
    \label{conv.ope}
    \end{center}
\end{figure}

The convolutional operation performed in the convolution layer can be expressed as
\begin{align}
s_{ijm}=\sum_{k=0}^{K-1}\sum_{p=0}^{H-1}\sum_{q=0}^{H-1}z_{i+p, j+q, k}W_{pqkm}+b_{m},
\end{align}
where $z_{ijk}$ is the input value at point $(i, j, k)$, $W_{pqkm}$ denotes the weight at point $(p, q, k)$ in the $m$-th filter, $b_m$ represents the bias of the $m$-th filter, and $s_{ijm}$ is the output of the convolution layer.  
The schematics of the convolutional operation and a convolution layer without bias are shown in figures \ref{conv.ope}($a$) and ($b$), respectively. 
The input is a three-dimensional matrix with the size of $L_1\times L_2\times K$, where $L_1$, $L_2$, and $K$ are the height, the width, and the number of channels (e.g., $K=3$ for RGB images), respectively.  
There are $M$ filters with the length $H$ and the $K$ channels. 
After passing the convolution layer, an activation function $f(\cdot)$ is applied to $s_{ijm}$, i.e.,
\begin{align}
z_{ijm}=f(s_{ijm}).
\end{align}
Usually, nonlinear {monotonic} functions are used as the activation function $f(\cdot)$.  
The sampling layer performs compression or extension procedures with respect to the input data.  
Here, we use a max pooling operation for the pooling layer, as summarized in \Figref{conv.ope}$(c)$. 
Through the max pooling operation, the machine learning model is able to obtain the robustness against rotation or translation of the images.  
In contrast, in the convolutional neural network autoencoder \cite{HS2006} (CNN-AE) explained below, the upsampling layer in the decoder part copies the values of the low-dimensional images into a high-dimensional field, {i.e., the nearest neighbour interpolation, as shown in figure \ref{conv.ope}$(d)$}.

The CNN-AE is composed of a CNN encoder ${\cal F}_e$, which maps high-dimensional data into a low-dimensional space, and a CNN decoder ${\cal F}_d$, which extends the data low-dimensionalized by the encoder part.  
If a CNN-AE ${\cal F}_c$ having a smaller latent vector $\tilde{\bm q}$ than the input $\bm q$ can generate the output identical to the input, it means that the dimension can be successfully reduced while retaining the original information.  
Summarizing above, the procedures of the CNN-AE are expressed as 
\begin{eqnarray}
{\bm q}_{\rm deco} \approx {\cal F}_c({{\bm q}}), ~~{\tilde{\bm q}} = {\cal F}_e({\bm q}), ~~{\bm q}_{\rm deco} = {\cal F}_d(\tilde{\bm q}){,}
\end{eqnarray}
{where $\vector{q}_{\textrm{deco}}$ denotes the decoder output.}

\begin{figure}[!t]
    \begin{center}
        \includegraphics[clip,width=0.8\textwidth ]{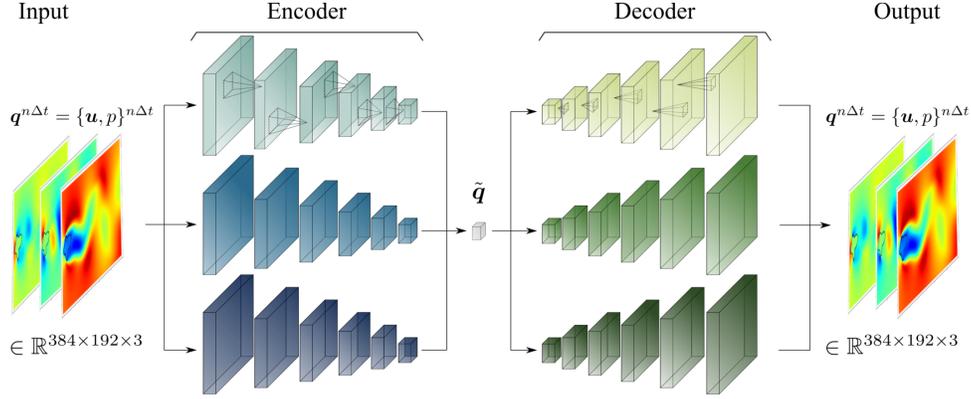}
    \caption{Schematic of the MS-CNN-AE. The layers represented by cubes in the encoder part include the convlutional layer, the batch normalization layer, the ReLU layer, and the max pooling layer in order. As for the decoder part, the cubes include the convolutional layer, the batch normalization layer, the ReLU layer, and the upsampling layer in order.}
    \label{CNN-AE_s}
    \end{center}
\end{figure}

In the present study, a multi-scale CNN-AE model (MS-CNN-AE) shown in figure \ref{CNN-AE_s} is proposed to reduce the spatial dimension of flow field data.
The MS-CNN-AE is inspired by the multi-scale CNN \cite{DQHG2018} developed for image-based super-resolution analysis to capture multi-scale sense of images.  
The size of three scales of filters are $3\times 3$, $5\times 5$, and $9\times 9$, respectively.  
As an example, the structure of the part to map the flow fields into the latent vector $\tilde{\bm q}\in{\mathbb R}^{6\times3\times4}$ (viz., the size of encoded values is $n_z=72$) is summarized in \Tabref{CNN-AE}.
There are batch normalization \cite{IS2015} layers between the convolution layer and the activation layer (ReLU) \cite{NH2010} to avoid the overfitting.
{The batch normalization, which normalizes the output of each unit based on the mean and variance in each training mini-batch, is known to accelerate learning by suppressing so-called internal covariate shift.}
The left and right parts of \Figref{CNN-AE_s} are the encoder and the decoder, respectively.  
The flow fields fed as the input are mapped by these three scales of filters, and then three encoded values $\in{\mathbb R}^{6\times3\times4}$ are obtained.  
These three encoded values are added in the {\it add layer} shown in \Tabref{CNN-AE}, and fed into {\it 7th Conv. layer} to obtain the encoded values representing the flow field in the low-dimensional space. Then, the decoder reconstructs the flow fields in the physical space from the encoded values {using upsampling layers}.

\begin{table}[!t]
\begin{center}
  \caption{Structure of each CNN-AE.}
  \label{CNN-AE}
  \small
  \begin{tabular}{ c c | c c} \hline
  \multicolumn{2}{c|}{Encoder} & \multicolumn{2}{c}{Decoder}\\\hline
  Layer&Output shape&Layer&Output shape\\ \hline\hline
  Input&(384, 192, 3)&8th Conv.&(6, 3, 4)\\ \hline
  1st Conv.&(384, 192, 16)&Batch Normalization&(6, 3, 4)\\ \hline
  Batch Normalization&(384, 192, 16)&ReLU&(6, 3, 4) \\ \hline
  ReLU&(384, 192, 16)&Dispersion Layer&(6, 3, 4)\\ \hline
  1st Max Pooling&(192, 96, 16)&1st Upsampling&(12, 6, 4)\\ \hline
  2nd Conv.&(192, 96, 8)&9th Conv.&(12, 6, 4)\\ \hline
  Batch Normalization&(192, 96, 8)&Batch Normalization&(12, 6, 4) \\ \hline
  ReLU&(192, 96, 8)&ReLU&(12, 6, 4) \\ \hline
  2nd Max Pooling&(96, 48, 8)&2nd Upsampling&(96, 48, 8)\\ \hline
  3rd Conv.&(96, 48, 8)&10th Conv.&(24, 12, 8)\\ \hline
  Batch Normalization&(96, 48, 8)&Batch Normalization&(24, 12, 8) \\ \hline
  ReLU&(96, 48, 8)&ReLU&(24, 12, 8) \\ \hline
  3rd Max Pooling&(48, 24, 8)&3rd Upsampling&(48, 24, 8)\\ \hline
  4th Conv.&(48, 24, 8)&11th Conv.&(48, 24, 8)\\ \hline
  Batch Normalization&(48, 24, 8)&Batch Normalization&(48, 24, 8) \\ \hline
  ReLU&(48, 24, 8)&ReLU&(48, 24, 8) \\ \hline
  4th Max Pooling&(24, 12, 8)&4th Upsampling&(96, 48, 8)\\ \hline
  5th Conv.&(24, 12, 8)&12th Conv.&(96, 48, 8)\\ \hline
  Batch Normalization&(24, 12, 8)&Batch Normalization&(96, 48, 8) \\ \hline
  ReLU&(24, 12, 8)&ReLU&(96, 48, 8) \\ \hline
  5th Max Pooling&(24, 12, 8)&5th Upsampling&(192, 96, 8)\\ \hline
  6th Conv.&(12, 6, 4)&13th Conv.&(192, 96, 16)\\ \hline
  Batch Normalization&(12, 6, 4)&Batch Normalization&(192, 96, 16) \\ \hline
  ReLU&(12, 6, 4)&ReLU&(192, 96, 16) \\ \hline
  6th Max Pooling&(6, 3, 4)&6th Upsampling&(384, 192, 16)\\ \hline
  Add Layer&(6, 3, 4)&14th Conv.&(384, 192, 3)\\ \hline
  7th Conv.&(6, 3, 4)&Batch Normalization&(384, 192, 3)\\ \hline
  Batch Normalization&(6, 3, 4)&ReLU&(384, 192, 3) \\ \hline
  ReLU&(6, 3, 4)&15th Conv. (Output)&(384, 192, 3) \\ \hline
      \end{tabular}
      \end{center}
\end{table}

Usually, the objective of regression tasks with supervised machine learning is to obtain optimized weights $\bm W$ by minimizing the predefined error function $\varepsilon$ such that ${\bm W}={\rm argmin}_{\bm W}||\varepsilon||_{\gamma}$, where $\gamma$ is the parameter of the norm.  
Here, we use a combination of the mean squared error $\varepsilon_m$ and the gradient difference loss $\varepsilon_g$ \cite{MC2016} as the loss function $\varepsilon$, i.e.,
\begin{align}
\varepsilon&=\varepsilon_m+\varepsilon_g,\\
\varepsilon_m&=\frac{1}{\hat{N_x}}\frac{1}{\hat{N_y}}\frac{1}{N_\phi}\sum_{i=1}^{\hat{N_x}}\sum_{j=1}^{\hat{N_y}}\sum_{k=1}^{N_\phi}(q_{(i,j,k)}-q_{\rm{deco} (\it{i,j,k})})^2,\\
\begin{split}
\varepsilon_g&=\frac{1}{\hat{N_x}}\frac{1}{\hat{N_y}}\frac{1}{N_\phi}\sum_{i=1}^{\hat{N_x}}\sum_{j=1}^{\hat{N_y}}\sum_{k=1}^{N_\phi}(|(q_{(i, j, k)}
-q_{(i-1, j, k)})-(q_{\textrm{deco} (i,j,k)}-q_{\textrm{deco} (i-1,j,k)})|\\
&\quad\quad\quad\quad\quad\quad\quad\quad\quad\quad+|(q_{(i, j-1, k)}-q_{(i, j, k)})-(q_{\textrm{deco} (i,j-1,k)}
-q_{\textrm{deco} (i,j,k)})|),
\end{split}
\end{align}
where
the subscripts represent the data indices.  
The gradient differential loss directly penalizes the gradient among grid points of the flow field data, and this feature enables the model to avoid blurry prediction \cite{SD2017}.  
Note that tuning of the weight between the mean squared error $\varepsilon_m$ and the gradient differential loss $\varepsilon_g$ is required, and its optimal weight varies depending on the problem.
In this study, the weight is set to $\varepsilon_m:\varepsilon_g=1:1$ following our preliminary test.  

The Adam algorithm \cite{KB2014} is applied as the optimizer for weight updating, and a four-fold cross validation is applied to train the models and avoid overfitting \cite{BK2019}.
\begin{figure}
    \begin{center}
        \includegraphics[clip,width=0.5\textwidth ]{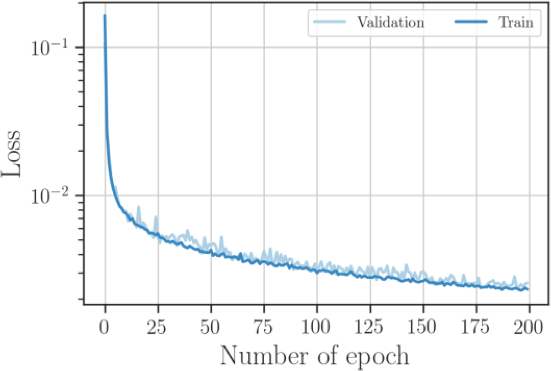}
    \caption{{An example of learning curve for the CNN-AE part.}}
    \label{fig_hist_cnn}
    \end{center}
\end{figure}
{
The minibatch size is set at 100 --- changing the minibatch size had no significant influence in our preliminary test.
The number of epochs is fixed at 200 (i.e., no early stopping).
Figure \ref{fig_hist_cnn} shows an example of the learning curve, which presents the relation between the number of epochs and the loss value.
The curve shows good convergence, and no overfitting is observed.
In the model evaluation, we use the best model which provides the lowest validation loss.
}

\subsubsection{Long Short-Term Memory (LSTM)}

\begin{figure}[b]
    \begin{center}
        \includegraphics[clip,width=0.7\textwidth]{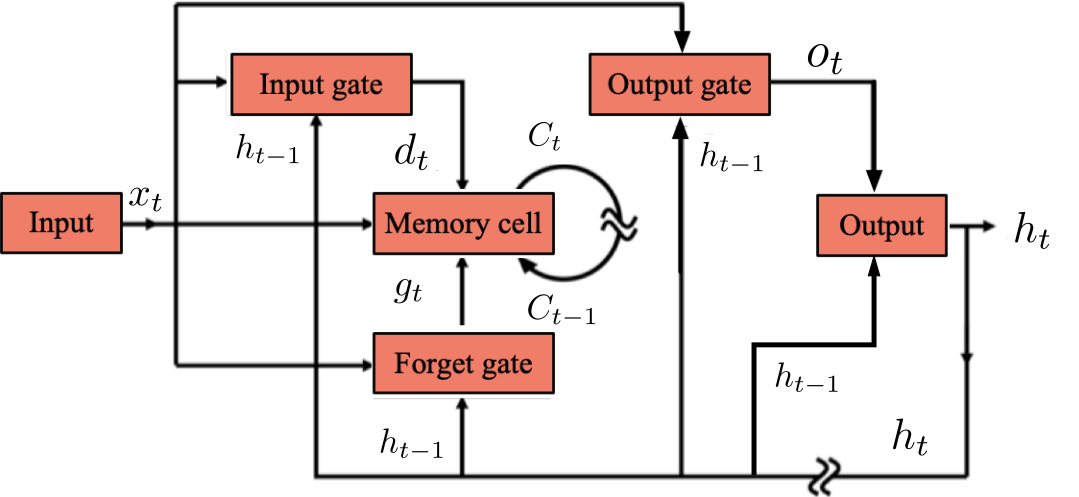}
    \caption{{Internal procedures of an LSTM.}}
    \label{fig5_20200121}
    \end{center}
\end{figure}
\begin{figure}[t]
    \begin{center}
        \includegraphics[clip,width=0.6\textwidth ]{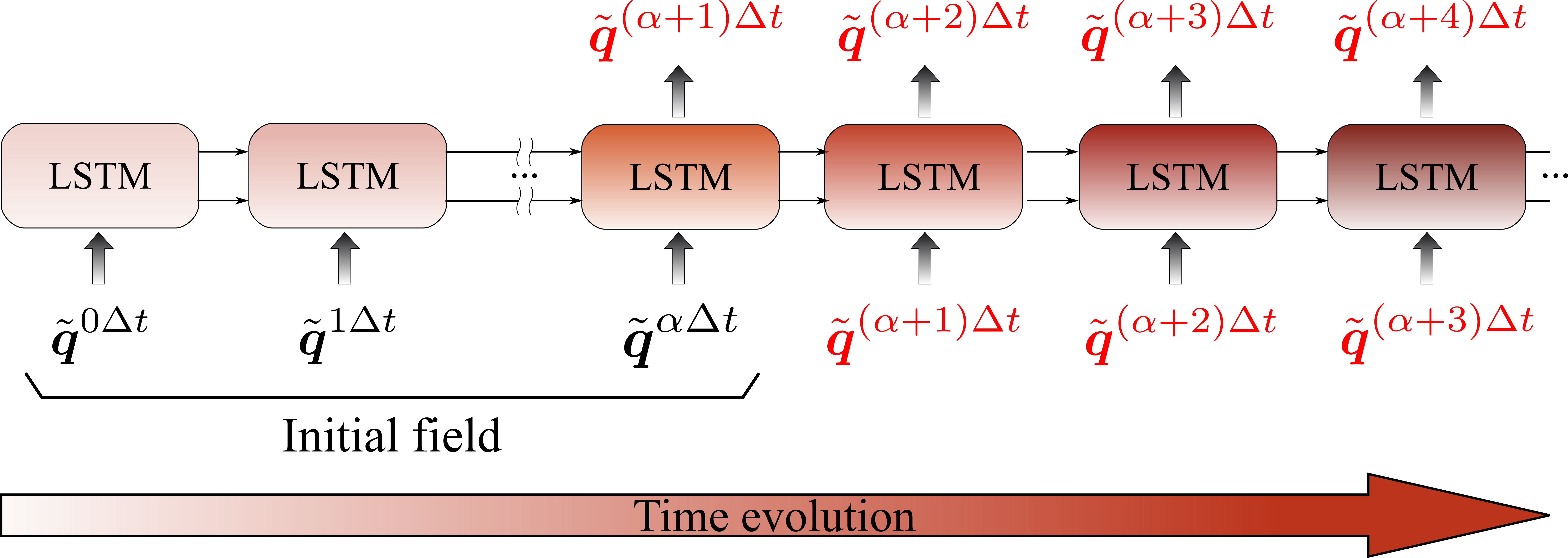}
    \caption{Schematic of the prediction {using} the LSTM {model}. The encoded values in black letters are the initial {fields} generated {by applying the CNN-AE to the DNS data}. The encoded values indicate the predicted fields by {the} LSTM from the previous outputs of {the} LSTM or the initial fields. The number of {the} initial fields is {$\alpha+1$} in th{is} figure.}
    \label{LSTM_s}
    \end{center}
    \begin{center}
        \includegraphics[clip,width=0.7\textwidth ]{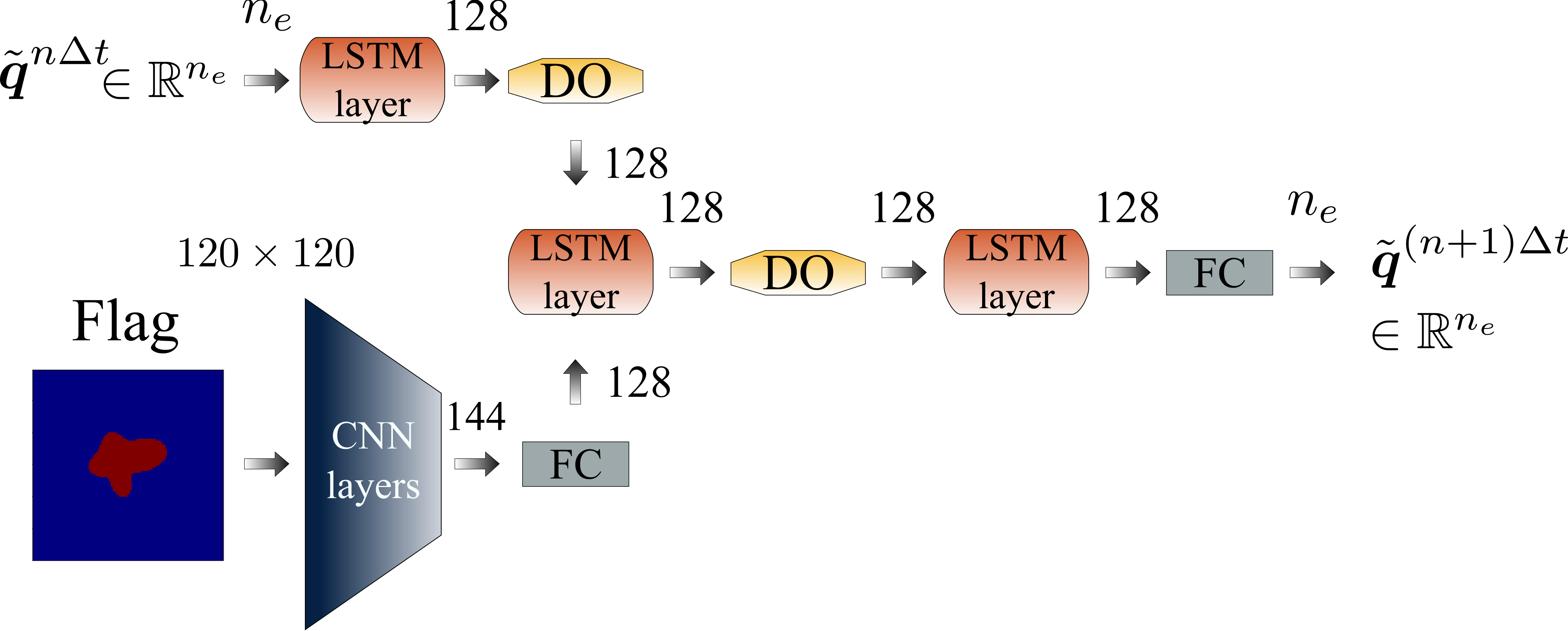}
    \caption{Schematics of the LSTM {model}. DO and FC in this figure represent dropout layers and fully-connected layers, respectively. The values {above} the arrows indicate the number of input/output of those layers, and $n_e$ represents the number of encoded values. Note that each LSTM layer has 128 units; viz., the output size of these layers is 128. }
    \label{LSTMs}
    \end{center}
\end{figure}

The long short-term memory (LSTM) \cite{HS1997} is a machine learning algorithm suited to handle {time-series problems}, e.g., speech recognition \cite{GJM2013}. 
The LSTM layer is composed of a cell, an input gate, an output gate, and a forget gate{, as illustrated in figure \ref{fig5_20200121}.}
The input gate is represented by $d$, output gate by $o$, and forget gate by $g$. 
The cell state is $C$ and the cell output is given by {$h_t$}, while the cell input is denoted as {$x_t$}, {where the subscripts represent a time step}. 
In sum, the internal procedures of the LSTM are formulated as
\begin{align}
d_t&=\sigma(W_d\cdot [h_{t-1}, x_t]+\beta_d),\\
o_t&=\sigma(W_o\cdot [h_{t-1}, x_t]+\beta_o),\\
g_t&=\sigma(W_g\cdot [h_{t-1}, x_t]+\beta_g),\\
\widetilde{C}_{t}&=\rm{tanh}(\it{W_c\cdot [h_{t-1}, x_t]+\beta_c}),\\
C_t&=g_t\times C_{t-1}+d_t\times\widetilde{C}_t,\\
h_t&=o_t\times\rm{tanh}(\it{C_t}),
\end{align}
where $W$ represents the weights for each gates and $\beta$ is the bias; the subscripts to $C$, $e$, and $h$ represent the time indices, and $\sigma$ is the sigmoid function.
Although readers are referred to literature \cite{HS1997} for further details, this structure enables the LSTM layer to deal with the time-series problem by keeping the previous input information in the cell state.

\begin{figure}[b]
    \begin{center}
        \includegraphics[clip,width=0.5\textwidth ]{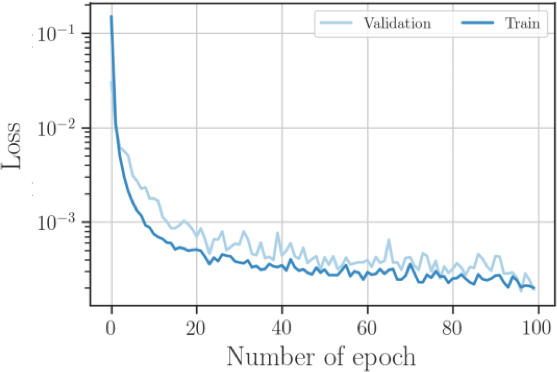}
    \caption{{An example of learning curve for the LSTM part.}}
    \label{fig_hist_lstm}
    \end{center}
\end{figure}

In this study, an LSTM model is employed to predict the temporal evolution of low-dimensionalized flow fields generated by the CNN-AE as illustrated in \Figref{LSTM_s}.  
In the diagram, $\tilde{\vector{q}}$ denotes the low-dimensional field, and the superscript represents time indices.  
The arbitrary number of the flow fields are fed into the LSTM model as the initial encoded fields.  
Next, the field predicted from these initial fields is recursively incorporated as the input data to the LSTM model keeping the cell state.  
The details of the present LSTM model are summarized in \Figref{LSTMs}.  
A dropout (DO) \cite{SHKSS2014} is applied in order to avoid overfitting.
A flag map of the bluff body (i.e., 1 for the bluff body region, 0 for the fluid region) is provided to the LSTM model as the information including the shape and boundary condition.  
Our preliminary test has shown that the model with the shape information outperforms the machine learning model without that information.

The mean squared error is used as the loss function $\tilde{\varepsilon}$ {to train} the LSTM model, i.e.,
$\tilde{\varepsilon}=\overline{(\tilde{\vector{q}}_{\rm{true}}-\tilde{\vector{q}}_{\rm{pred}})^2}$,
where $\tilde{\vector{q}}_{\rm{true}}$ is the true encoded field, $\tilde{\vector{q}}_{\rm{pred}}$ is the field predicted by the LSTM model, and the overbar represents the average similar to equation (9).
{
The solution data set is prepared from the output of the CNN-AE, and the LSTM model is trained using {\it teacher forcing} \cite{WZ1989}.
}
{Following our preliminary test, the number of time sequences used for the training process is set to 20.
Hence, the training for the LSTM model is equivalent to optimizing the weights in the LSTM model ${\bm w}_L$ such that
\begin{eqnarray}
{\bm w}_L = {\rm argmin}_{{\bm w}_L}||{\tilde{\bm q}}^{(n+1)\Delta t} - {\cal F}_L({\tilde{\bm q}}^{n\Delta t},{\tilde{\bm q}}^{(n-1)\Delta t},{\tilde{\bm q}}^{(n-2)\Delta t},...,{\tilde{\bm q}}^{(n-19)\Delta t})||_2,
\end{eqnarray}
where the subscript of ``true'' is omitted for brevity.
}
Similarly to the CNN-AE above, the Adam algorithm \cite{KB2014} is applied as the optimizer, a four-fold cross validation is used, {and the best model which provides the lowest validation loss in the learning process is used for the model evaluation.
Both the minibatch size and the number of epochs are set to 100.
An example of learning curve for the LSTM part is presented in figure \ref{fig_hist_lstm}, which shows good convergence and no overfitting.}

{For the model evaluation,} the number of time steps used for the input to the LSTM ${\cal F}_L$ is set to 1 such that ${\tilde{\bm q}}^{(n+1)\Delta t}={\cal F}_L({\tilde{\bm q}}^{n\Delta t})$ except for the first iteration.
For the first iteration, the {latent vector at the} next time step is obtained from {the solution data of the 5 initial time steps (i.e., $\alpha=4$ in figure 7).}
In sum, the temporal evolution of the mapped vector in the LSTM is formulated as
\begin{eqnarray}
{\tilde{\bm q}}^{5\Delta t}={\cal F}_L(\tilde{\bm q}^{4\Delta t},\tilde{\bm q}^{3\Delta t},\tilde{\bm q}^{2\Delta t},\tilde{\bm q}^{1\Delta t},\tilde{\bm q}^{0\Delta t}),\\
\tilde{\bm q}^{(n+1)\Delta t}={\cal F}_L(\tilde{\bm q}^{n\Delta t}),~~n\geq 5.
\end{eqnarray}
Note that our preliminary test has shown that the results are not sensitive to the number of time steps used for the first iteration.

\subsubsection{Machine-learning based reduced order model (ML-ROM)}

\begin{figure}[b]
    \begin{center}
        \includegraphics[clip,width=0.6\textwidth ]{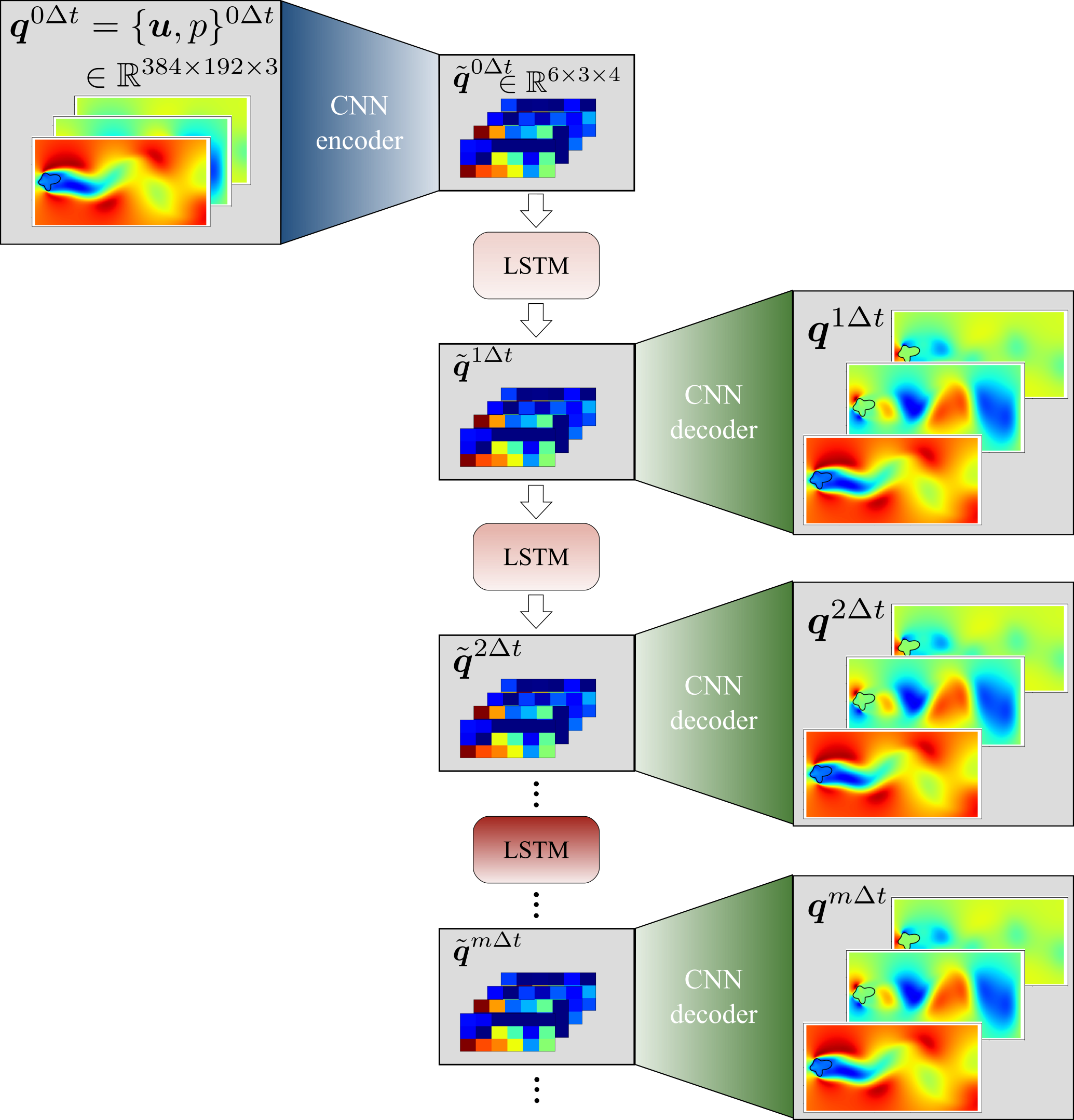}
    \caption{Schematic of the ML-ROM with the latent space size of $6\times3\times4$. The number of time steps of the initial field is set to 1 for illustration purpose. The compressed vector obtained by using the CNN encoder evolves temporally using LSTM. The temporal evolution of the flow field is recovered by using the CNN decoder.}
    \label{ML-ROM}
    \end{center}
\end{figure}

As illustrated in \Figref{ML-ROM}, the proposed machine-learning based reduced order model (ML-ROM) is a combination of the MS-CNN-AE model and the LSTM model introduced above. 
The initial flow fields generated by DNS are fed into the trained CNN encoder to map those into the latent space.
By feeding the obtained latent vectors to the trained LSTM model, it predicts the latent vector at the next time step.  
The LSTM model recursively predicts the temporal evolution of the encoded fields by using the previous output as the input.  
The temporal evolution of the flow field in the physical space can be recovered by using the trained CNN decoder.
Note that the number of initial flow fields in this figure is set to 1 for simplicity of illustration.

\section{Results and Discussion}

\subsection{Assessment of ML-ROM for wakes behind various random shapes}

\begin{figure}[b!]
    \begin{center}
        \includegraphics[clip,width=0.8\textwidth ]{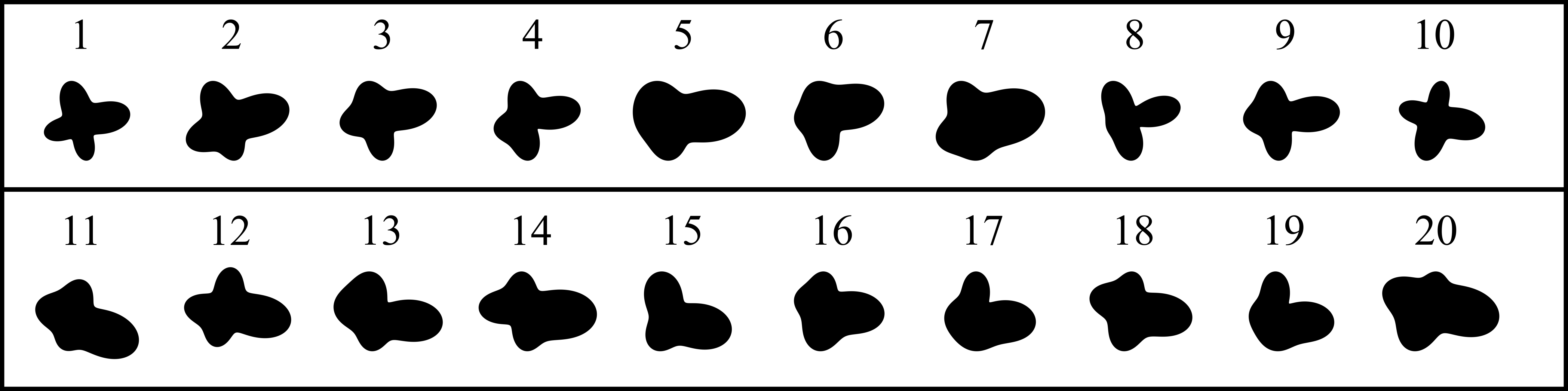}
    \caption{The bluff body shapes of the test data set used to evaluate the machine learning models. The number shown above each shape represents the shape number.}
    \label{shape}
    \end{center}
    \begin{center}
        \includegraphics[clip,width=0.9\textwidth ]{fig12.png}
    \caption{Instantaneous flow fields for various bluff bodies. Flow fields computed by the DNS and those reconstructed by the MS-CNN-AE model are compared.}
    \label{VSB_CNN_flows}
    \vspace{-3mm}
    \end{center}
\end{figure}

As a proof of concept to establish an ML-ROM for unseen data, we use the data sets of bluff bodies with various random shapes, as explained in section 2.  
In this subsection, the MS-CNN-AE is developed first to map the high dimensional flow ${\mathbb R}^{\in 384\times192\times3}$ into a latent space ${\mathbb R}^{\in 6\times3\times4}$.
Then, the LSTM part is trained to learn the temporal evolution of the obtained latent vectors.
Note that the dependence on {the latent vector size} will be examined in the next subsection. 

The MS-CNN-AE is trained by using the data set which consists of flow data for 80 different bluff bodies with the 500 instantaneous time-series fields prepared for each bluff body shape.  
This model is evaluated by the test data set, which are different from those used for training.  
The test data set includes flows around bluff bodies for 20 different shapes shown in \Figref{shape}.  


\begin{figure}[!t]
    \begin{center}
        \includegraphics[clip,width=0.9\textwidth ]{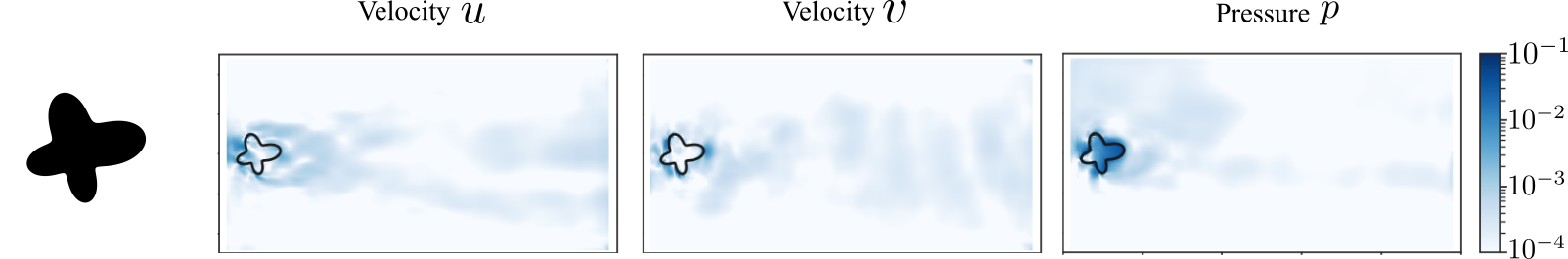}
    \caption{{Time-averaged local squared error fields of MS-CNN-AE for shape number 1.}}
    \label{cnn_err}
    \vspace{-3mm}
    \end{center}
\end{figure}

The flow fields computed by the DNS and those reconstructed by the MS-CNN-AE are summarized in \Figref{VSB_CNN_flows}.  
In this figure, the flows with shape numbers of 1, 3, 5, 7, 9, 11, 13 and 15 are shown as the examples.  
The reconstructed flow fields show good agreement with the reference DNS fields.
Although not shown here, the results with other bluff body shapes have similar trends to \Figref{VSB_CNN_flows}.
{Time-averaged local squared error fields for shape number 1 are shown in figure \ref{cnn_err}.
Although the error is concentrated near the bluff body, the error is sufficiently small in the wake region. 
}

\begin{figure}
    \begin{center}
        \includegraphics[clip,width=0.9\textwidth ]{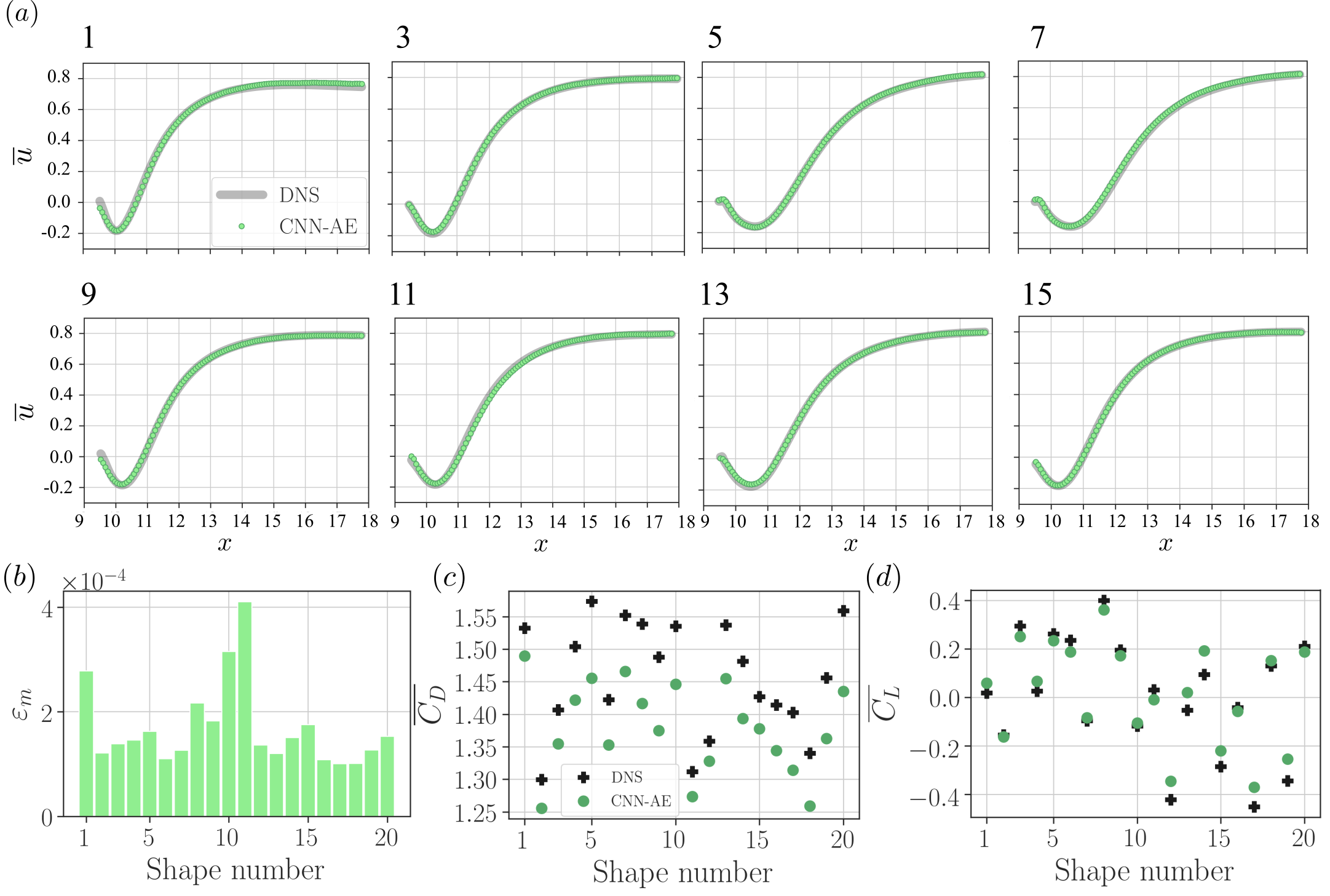}
    \caption{Assessments of the MS-CNN-AE model for flows around various bluff bodies at ${\rm Re}_D=100$:  $(a)$ {m}ean streamwise velocities on the centerline for shape numbers 1, 3, 5, 7, 9, 11, 13 and 15; $(b)$ {m}ean squared errors against the reference DNS data; $(c)$ {t}ime-averaged drag coefficient; $(d)$ {t}ime-averaged lift coefficient.}
    \label{VSB_CNN_base}
    \end{center}
\end{figure}

The mean streamwise velocities on the centerline of the wake are presented in \Figref{VSB_CNN_base}$(a)$.  
The reconstructed centerline velocities are in excellent agreement with the reference DNS data.  
The mean squared errors, the time-averaged drag and lift coefficients are also summarized in figures~\ref{VSB_CNN_base}$(b), (c)$ and $(d)$, which indicate that the mean squared errors are sufficiently small and the averaged force coefficients of the reconstructed fields reasonably match the DNS values.

\begin{figure}
    \begin{center}
        \includegraphics[clip,width=0.9\textwidth ]{fig15.png}
    \caption{Instantaneous flow fields with various bluff bodies { at $t=25$}. The DNS and reconstructed flow fields by the ML-ROM are summarized.}
    \label{VSB_LSTM_flows}
    \end{center}
    \begin{center}
        \includegraphics[clip,width=0.9\textwidth ]{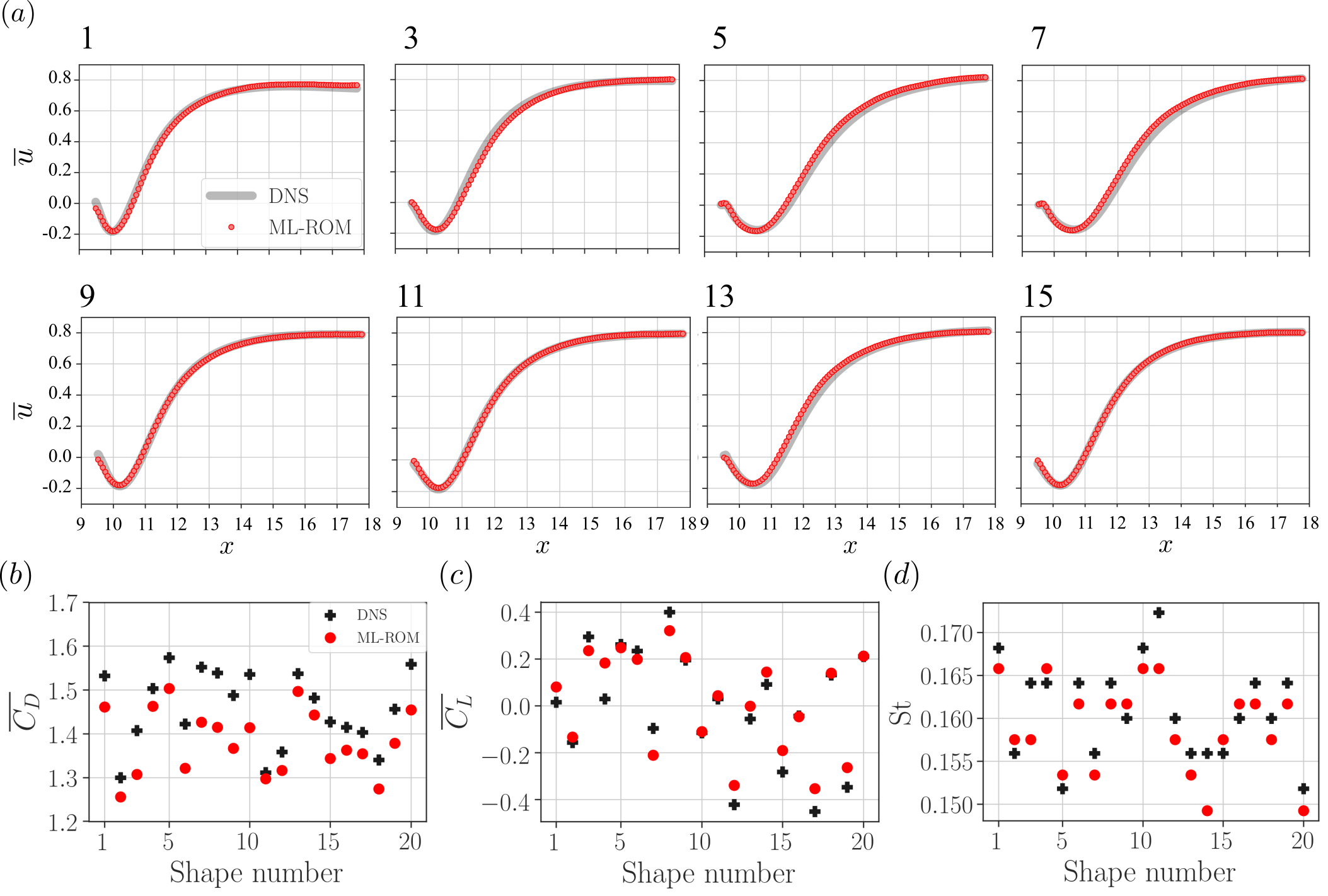}
    \caption{Assessments of ML-ROM with various shapes wake at ${\rm Re}_D=100$:  $(a)$ {m}ean streamwise velocit{y} on the centerline for shape numbers 1, 3, 5, 7, 9, 11, 13 and 15; $(b)$ {d}rag coefficient; $(c)$ {l}ift coefficient; $(d)$ Strouhal number.}
    \label{VSB_LSTM_base}
    \end{center}
\end{figure}

The LSTM is trained by using the time step of $\Delta t=0.25$ to learn the temporal evolution of the low-dimensionalized fields for the 80 different bluff bodies obtained by the MS-CNN-AE to construct the ML-ROM, as illustrated in \Figref{ML-ROM}.  
The amount of the training and validation data is 40000, which consist of 500 time-series data for each bluff body.  
Five instantaneous flows are prepared for each shape as the initial fields of the predictions, as mentioned above. 
Some instantaneous fields predicted by the ML-ROM {after 100 recursive iterations corresponding to $t=25$} are compared to the DNS data in \Figref{VSB_LSTM_flows}.
{Both} flows are observed to be similar for all attributes.  

The statistical assessments of the prediction by the ML-ROM are summarized in \Figref{VSB_LSTM_base}.  
The predicted results are again in good agreement with the reference DNS data in terms of the mean centerline velocity and the force coefficients, which suggests that the present ML-ROM can successfully capture the feature of the unsteady wake.
As shown in figure \ref{VSB_LSTM_base}$(d)$, the Strouhal number $\rm St$ is also well predicted, which confirms that the temporal structure is also well reproduced by the LSTM part even for the flows not used for the training
(note again that shapes 1--20 are not used in the training process).

\begin{figure}
    \begin{center}
        \includegraphics[clip,width=0.9\textwidth ]{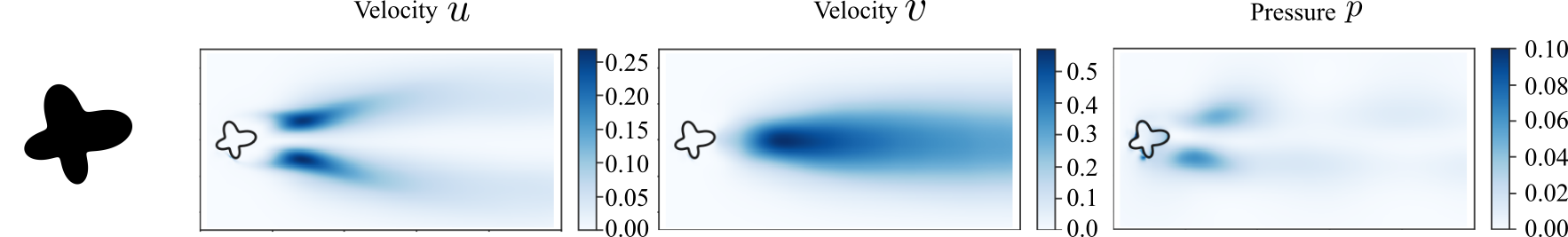}
    \caption{{Time-averaged local squared error fields of ML-ROM for shape number 1.}}
    \label{lstm_err}
    \vspace{-3mm}
    \end{center}
\end{figure}
\begin{figure}
    \begin{center}
        \includegraphics[clip,width=0.4\textwidth ]{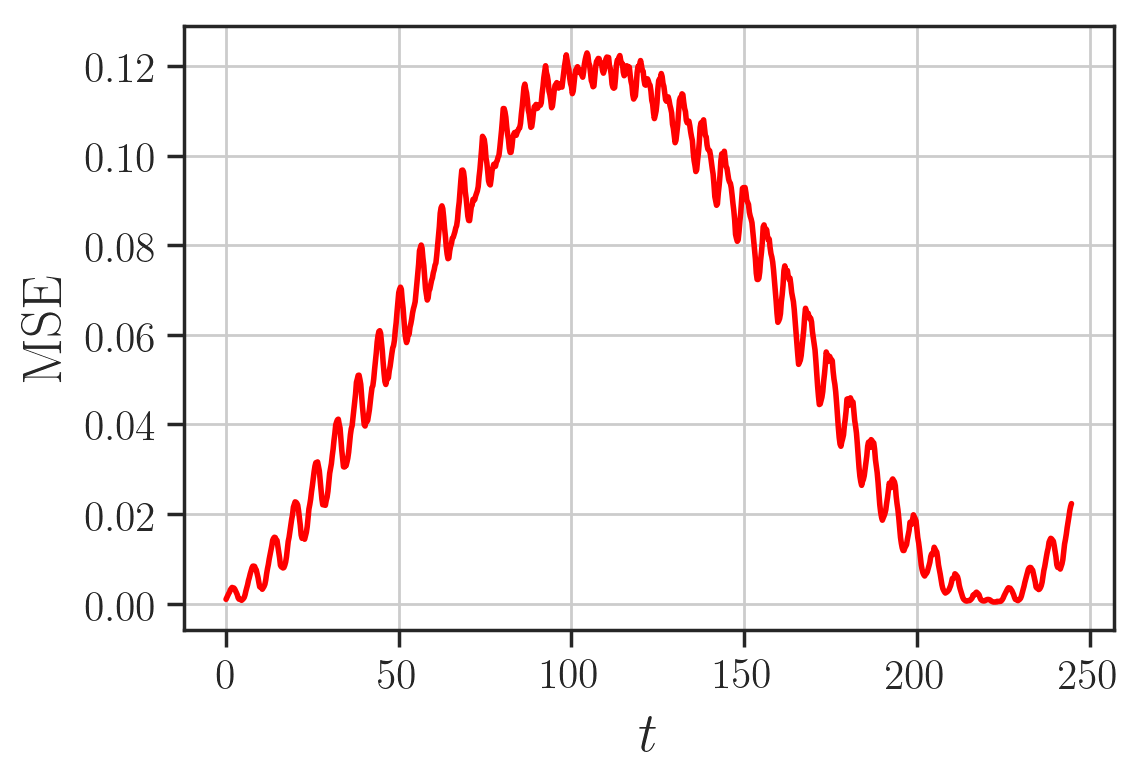}
    \caption{{Time trace of mean squared error of ML-ROM for shape number 1.}}
    \label{err-time-ML-ROM}
    \end{center}
\end{figure}

{We also present in figure \ref{lstm_err} the time-averaged local squared error computed using 1000 recursive inputs.
Because of the recursive input, the time-averaged error is concentrated in the wake region, especially where the fluctuations are large.
The time trace of the mean squared error is also shown in figure \ref{err-time-ML-ROM}.
The error varies periodically in time due to the small difference in the Strouhal number (figure \ref{VSB_LSTM_base}$(d)$), but it does not grow.}
Summarizing above, the present ML-ROM is confirmed to have the ability to predict the flows around various bluff bodies.

\subsection{Influence of the parameters}

In the aforementioned discussion, we have set the size of the latent vector in the MS-CNN-AE to be $n_z=72~(=6\times3\times4)$ and the time steps in between the mapped vectors for the LSTM to be $\Delta t=0.25$.
In this subsection, we discuss the influence of these parameters.

\subsubsection{Dependence on the latent {vector} size in the MS-CNN-AE}

\begin{figure}
    \begin{center}
        \includegraphics[clip,width=0.9\textwidth]{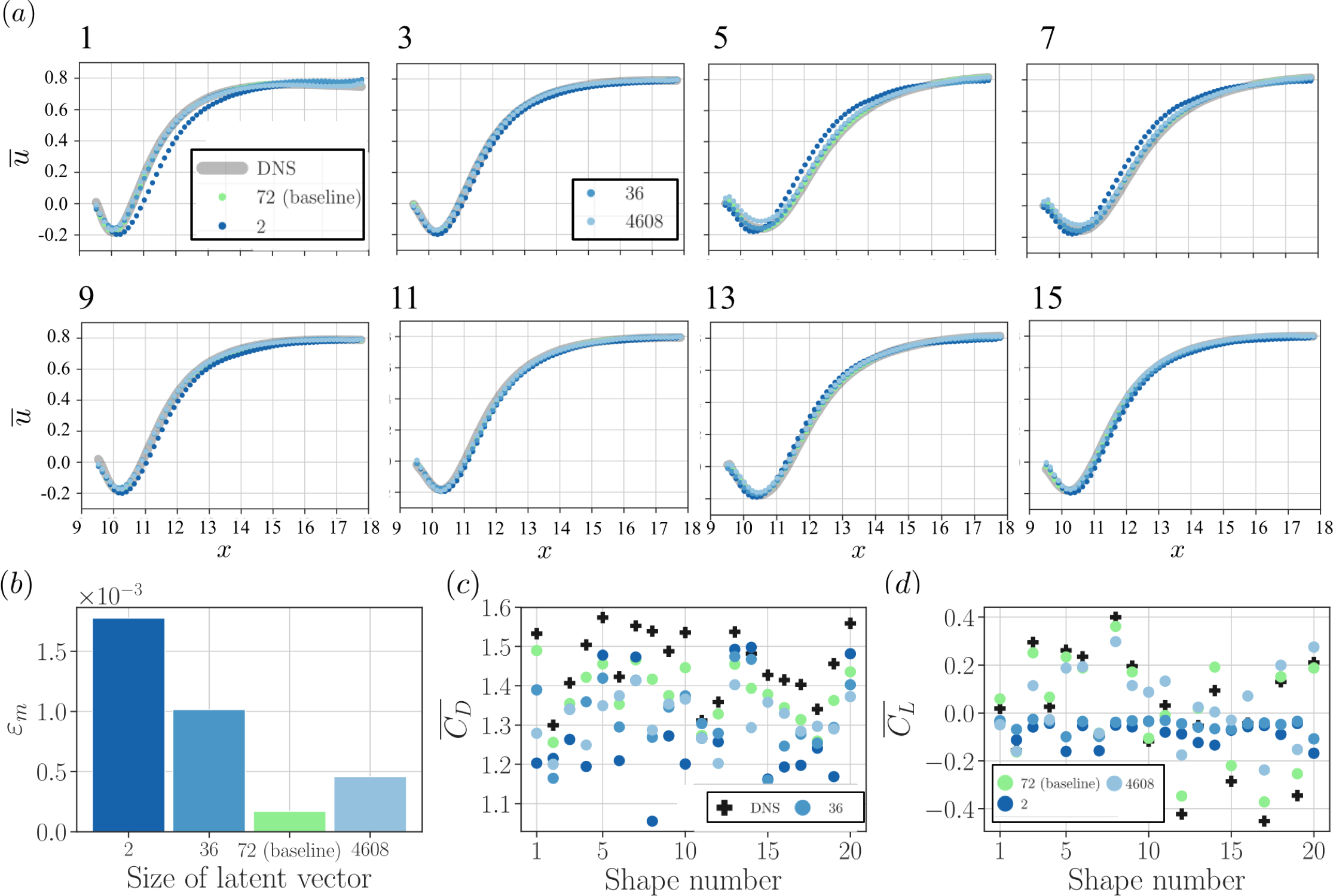}
    \caption{Dependence on the latent vector size in the MS-CNN-AE:  $(a)$ {m}ean streamwise velocity on the centerline; $(b)$ {m}ean squared error; $(c)$ {t}ime-averaged drag coefficient; $(d)$ {t}ime-averaged lift coefficient.  Here, four-fold cross validation are arranged.}
    \label{fig13}
    \end{center}
\end{figure}

The dependence on the latent vector {size} $n_z$ in the MS-CNN-AE is investigated and summarized in \Figref{fig13}.  
Here, we examine $n_z=2$, 36, 72 (baseline), and 4608.  
Since the temporal evolution of the mapped vector is obtained by the LSTM, which has a fully-connected structure between layers, a smaller latent vector allows us to establish an ML-ROM at a lower computational cost.

As shown in figure \ref{fig13}$(a)$, the mean centerline velocity looks reasonably well reproduced in all cases.  
However, the mean velocities for some shapes, i.e., shapes 1, 5, and 7, are underestimated with $n_z=2$ and 36.  
Similar trends can also be seen in the assessment of the force coefficients as summarized in figures \ref{fig13}$(c)$ and $(d)$.  
It suggests that $n_z=72$ is the minimum size required to reconstruct the present flow fields with an appropriate fidelity. 
It is also surprising that the error $\varepsilon$ with $n_z=72$ is smaller than that with $n_z=4608$, as shown in figure \ref{fig13}$(b)$.
This is likely due to the structure of the CNN-AE, which has more pooling operations for $n_z=72$ case.
It is widely known that incorporating the pooling operation in CNN structures enables the models to retain the robustness against the rotation or translation of the images because the sensitivity is decreased \cite{LBBH1998}.
It indicates that the model with $n_z=72$ is better than that with $n_z=4608$ in terms of generality for unknown wakes thanks to the aforementioned robustness, especially in the present case where the wakes behind random shapes are considered.

Summarizing above, over-compression of the input and output flow data has a risk to lack the spatially coherent information of the flow field because of the pooling operation; however, the appropriate number of the pooling operation allows us to keep the robustness for unseen data.

\subsubsection{Dependence on the time step {size} in the LSTM}

 \begin{figure}
     \begin{center}
         \includegraphics[clip,width=0.9\textwidth ]{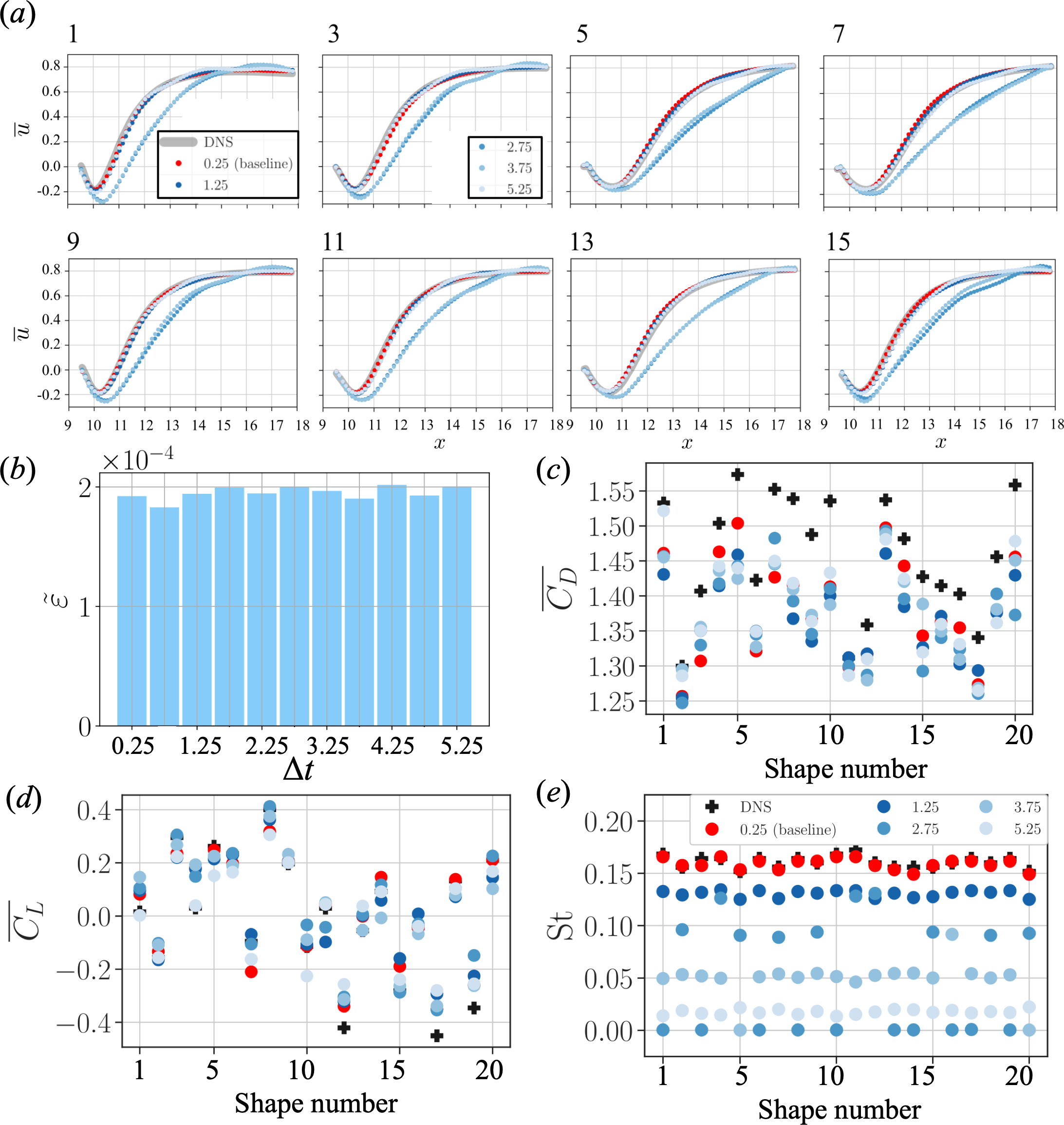}     
         \caption{Dependence on the time step {size} in the LSTM{:}  $(a)$ {m}ean streamwise velocity on the centerline{;}  $(b)$ relationship between time step and the $L_2$ error{;} $(c)$ {t}ime-averaged drag coefficient{;} $(d)$ {t}ime-averaged lift coefficient{;} $(e)$ Strouhal number. Here, four-fold cross validation are arranged.}
    \label{fig14}
     \end{center}
 \end{figure}

For high-fidelity simulations such as DNS and large eddy simulation, the time step size is always limited by numerical constraints.
Thus, it would be attractive if the present ML-ROM can be used with substantially wider time step sizes.

Let us examine the dependence on the time step size in the LSTM, as summarized in figure \ref{fig14}.  
Here, we consider 11 cases: $\Delta t= 0.25$ (baseline)--5.25 with an increment of 0.50 in dimensionless time, although only the cases with $\Delta t=0.25$, 1.25, 2.75, 3.75, and 5.25 are shown in figures \ref{fig14}$(a)$, $(c)$, $(d)$, and $(e)$.
{Recall that the time step used in the DNS was $\Delta t=2.5\times10^{-3}$; namely, the baseline time step  of $\Delta t= 0.25$ used for the LSTM is already 100 times wider than that.}
As shown here, the basic trend observed for the all assessments is that the error increases with the time step size, especially for $\Delta t=2.75$ and 3.75.

It is worth noting that the mean centerline velocity profile and the force coefficients are in reasonable agreement even with $\Delta t =5.25$.
However, in this case, the ML-ROM is considered to learn a typical aliasing signal because the sampling interval $\Delta t =5.25$ is close to one period of the actual flow $T\simeq 6$.  
The Strouhal number predicted with $\Delta t = 5.25$ are around ${\rm St}\simeq 0.02$  for all Reynolds numbers  as shown in figure \ref{fig14}$(e)$, which is also consistent with the value for the $-1$ aliasing at this sampling rate (i.e. $|1/T - 1/\Delta t|\simeq 0.02$).
A similar argument holds for the cases of $\Delta t =3.75$ or 2.75 where the sampling interval is longer than or just around the interval corresponding to the Nyquist frequency of the present periodic signal.

We note in passing that the results of the present ML-ROM also depends on the number of time steps used for the input of LSTM to predict the field of the next time step.
We used 20 time steps for the training process of LSTM, but significant dependence was not observed in our preliminary test as far as more than 5 time steps were used. 
This is likely due to the periodic nature of this specific flow.
Otherwise, the number of input time steps used for the training process is also {a} crucial factor, and it should be chosen depending on user's requirements.

\section{Conclusions}

We presented machine-learning-based reduced order modeling for unsteady flows.
A convolutional neural network based autoencoder (CNN-AE) was employed to map a high-dimensional flow field into a low-dimensional latent space, and a long short term memory (LSTM) was utilized to deal with the temporal evolution of the low-dimensionalized vectors obtained by the CNN-AE.
As a test case, flows around bluff bodies with various shapes were considered.  
The flows predicted by the machine-learned reduced order model (ML-ROM) showed statistically good agreement with the reference DNS data also for unseen bluff body shapes not used in the training process, which suggests that the present ML-ROM learns not just the flow fields used for training but the physics governed by the Navier--Stokes equation under different geometrical configurations.

Moreover, some case studies were conducted to investigate the dependence on the parameters used for the ML-ROM.  
The size of the latent vector of the CNN-AE model has relatively small influence on the reconstruction ability, but this might be specific to the present problem with temporal periodicity.  
We also found that the structure of the CNN-AE allows us to keep the robustness for unseen flow data.  
Concerning the dependence on the time step size used in the LSTM, the error increased with the time step size between the mapped vectors.
The value of $\Delta t=0.25$, which corresponds to about 20 subdivision of one period of vortex shedding, can be recommended from the present study to reproduce the Strouhal number accurately.

The present study was a proof of concept to establish an ML-ROM for more general fluid dynamics. 
{It should be stressed again{, however,} that the present proof of concept was performed with a limited range of shapes, and that more variability will be required in practice.}
Although laminar periodic flows are considered as the problem setting in the present study, the proposed idea can be further extended to more complex phenomena, e.g., three dimensional flows at high Reynolds numbers.
{Concerning the possibility of applying LSTM to turbulent flows, Srinivasan et al. \cite{SGASR2019} have recently demonstrated that the chaotic temporal evolution of the nine-equation turbulent shear flow model can be well captured by utilizing the LSTM, as mentioned in the introduction.
Therefore, the key issue for the present type of ML-ROM to be applied to more complex flows should be the development of a more efficient --- and preferably interpretable --- low-dimensionalization method, as is tackled by different research groups \cite{MFF2020,ENMN2019}. }

\begin{acknowledgements}
Authors are grateful to Dr.~S.~Obi, Dr.~K.~Ando, Mr. M.~Morimoto and Mr. T.~Nakamura (Keio University) for fruitful discussions. 
This work was supported through JSPS KAKENHI Grant Number 18H03758 by Japan Society for the Promotion of Science.
\end{acknowledgements}

\end{document}